\documentclass[]{elsart}

\usepackage{graphics}
\usepackage{graphicx}
\usepackage{epsfig}

\usepackage{amssymb}

\begin{document}

\begin{frontmatter}



\title{Early nonlinear regime of MHD internal modes: the resistive case}


\author{M.C. Firpo}
%\ead{firpo@lptp.polytechnique.fr}
\address{Laboratoire de Physique et Technologie des Plasmas (C.N.R.S. UMR 7648),
Ecole Polytechnique, 91128 Palaiseau cedex, France}
%\thanks[]{Physics Letters A \textbf{342} (2005) 263-266.}

%\thanks[]{Expanded version of a talk given at the 9th Plasma Easter Meeting, March 30th - April 1st 2005, Villa Gualino, Turin}
%\thanks[]{Physics Letters A \textbf{342} (2005) 263-266.}

\begin{abstract}
It is shown that the critical layer analysis, involved in the linear
theory of internal modes, can be extended continuously into the
early nonlinear regime. For the $m=1$ resistive mode, the dynamical
analysis involves two small parameters: the inverse of the magnetic
Reynolds number $S$ and the $m=1$ mode amplitude $A$, that measures
the amount of nonlinearities in the system. The location of the
instantaneous critical layer and the dominant dynamical equations
inside it are evaluated self-consistently, as $A$ increases and
crosses some $S$-dependent thresholds. A special emphasis is put on
the influence of the initial $q$-profile on the early nonlinear
behavior. Predictions are given for a family of $q$-profiles,
including the important low shear case, and shown to be consistent
with recent experimental observations.
\end{abstract}

\end{frontmatter}

% main text

The $m=n=1$ internal modes, such that the safety factor goes below one for
some inner radius, remain critical macroscopic modes for large scale tokamak
plasma dynamics and confinement. They are particularly involved in sawtooth
oscillations and crashes. This is a common deleterious phenomenon as
conventional tokamak discharges eventually operate with $q_{0}<1$ since
current density tends to a peaked profile. Additionally, the $m=n=1$
internal modes form a laboratory prototype for reconnection. Such phenomena
typically proceed \textit{beyond linear regime}.

We shall consider here the $m=n=1$ purely resistive mode \cite{Coppi76} that
is ideally marginally stable. The original motivation of this work was to
understand the growth of the $m=1$ resistive mode up to its nonlinear
saturation, on the basis of some striking numerical simulations performed by
Aydemir \cite{Aydemir97} and previous observations \cite{waddell}. Within
the reduced MHD framework in cylindrical coordinates and some given $q$%
-profile \cite{Aydemir97}, the time behavior of the kinetic energy in the $%
m=1$ mode amounts to an initial exponential growth consistent with the
linear regime, followed by a transient stage where the growth rate
decreases, that is brutally interrupted by a second exponential growth in
the nonlinear regime. This second exponential stage eventually terminates,
as the kinetic energy in the $m=1$ mode saturates which coincides with the
completion of magnetic reconnection.

The reduced MHD system under consideration reads
\begin{eqnarray}
\frac{\partial U}{\partial t} &=&\left[ \phi ,U\right] +\left[ J,\psi \right]%
,  \label{finsys_1} \\
\frac{\partial \psi }{\partial t} &=&\left[ \phi ,\psi \right] +\eta
(J-J_{0}).  \label{finsys_2}
\end{eqnarray}
Helical symmetry is assumed: The poloidal and toroidal angles, respectively $%
\theta$ and $\varphi$, only come in through the helical angle $\alpha =
\varphi - \theta$. $\phi$ and $\psi$ are the plasma velocity and helical
magnetic field potentials: the velocity is $\mathbf{v}=\mathbf{\hat{\varphi}}%
\times \mathbf{\nabla }_{\bot }\phi $ and the magnetic field $\mathbf{B}%
=B_{0\varphi }\mathbf{\hat{\varphi}}+\mathbf{\hat{\varphi}}\times \mathbf{%
\nabla }_{\bot }\left( \psi -r^{2}/2\right)$. $U=\nabla _{\bot }^{2}\phi $
is the vorticity and $J=\nabla _{\bot }^{2}\psi $ the helical current
density, with $\nabla _{\bot }^{2}\equiv r^{-1}\partial _{r}r\partial
_{r}+r^{-2}\partial _{\alpha }^{2}$. Poisson brackets are defined by $\left[
\phi ,U\right] =-\mathbf{\hat{\varphi}}\cdot ( \mathbf{\nabla }_{\bot }\phi
\times \mathbf{\nabla }_{\bot }U) =r^{-1}( \partial _{r}\phi \partial
_{\alpha }U-\partial _{r}U\partial _{\alpha }\phi)$. Eqs. (\ref{finsys_1})-(%
\ref{finsys_2}) are dimensionless: Time has been normalized to the poloidal
Alfv\'{e}n time, the radial variable $r$ to the minor radius, and $\eta$ is
the inverse of the magnetic Reynolds number $S$, and is given by the ratio
of the poloidal Alv\'{e}n time to the resistive one. In high-temperature
fusion plasmas, $\eta$ is typically much smaller than one.

Consider equilibria such that, for some internal radius $r_{s0}<1$, $%
q(r_{s0})=1$, that is $\psi _{0}^{\prime }(r_{s0})=0$. Then, due to the
Ohm's law (\ref{finsys_2}), plasma volume divides in two region. Far from
the $q=1$ surface (outer domain), plasma behaves ideally whereas, in the
vicinity of $r_{s0}$ (inner region), resistivity plays a crucial,
destabilizing, role. Linear theory \cite{Coppi76} uses asymptotic matching
analysis to provide $m=1$ eigenfunctions in the form $A(t)f_{L}(r)\exp(i%
\alpha)$ valid in the whole domain. In the outer (ideal) domain, this
solution is valid, that is nonlinear effects are negligible, as long as $A
\ll 1$ \cite{Firpo04}. Injecting the linear solutions $\psi_{1}(r,%
\alpha,t)=A(t)\psi_{L}(r)\exp(i\alpha)$ and $\phi_{1}(r,\alpha,t)=A(t)%
\phi_{L}(r)\exp(i\alpha)$ into (\ref{finsys_1})-(\ref{finsys_2}) calls for
an amplitude expansion. The procedure has been given in Refs. \cite%
{Firpo04,FirpoCoppi03}. The particularity of the linear radial
eigenfunctions $\psi_{L}(r)$ and $\phi_{L}(r)$, that needs a careful
consideration, is that they have strong gradients inside the critical layer.
More precisely, their radial derivatives are of the order of the inverse of
the critical layer width, that is $\mathcal{O}(\eta^{-1/3})$. This means in
particular that this approach restricts to situations strictly above
marginal stability and where the linear regime is well defined, with clear
scalings, yielding the resistive ordering, and non-pathological $q$-profiles
(in the sense of Ref. \cite{Fitzpatrick1989}).

We wish then to answer the question: ``How does the $m=1$ resistive mode
develop into the nonlinear regime ?" To do this, let us first recognize that
the problem involves \textit{two} small parameters. An obvious one is the
resistivity $\eta$. However, considering it to be the only one small
parameter, in some perturbation analysis with conventional expansions of the
type $f=f_{0}+\eta f_{1} + \ldots$ would lead to a dead end: this would
bring up a singular expansion, with additional $\eta \ln(\eta)$ terms, with
no asymptotic validity unless assuming that the mode amplitude is always
kept vanishingly small. It is interesting to note that such a procedure
would actually be valid for the tearing mode with the small parameter limit $%
\Delta ^{\prime}$ \cite{EscOtt2004,MilPorc04}. In the present case, such a
perturbation analysis would be ill-posed. A second small parameter enters
the game, the $m=1$ mode amplitude $A$ that can be viewed as an indicator of
the amount of nonlinearities in the system. As previously said, the approach
will then be that of an amplitude expansion.

The first step will be to determine the end of validity of the
linear regime. In the outer domain, this occurs for $A$ of order one
\cite{Firpo04} but, in the inner domain, the linear solution breaks
earlier. This occurs when mode coupling terms such as $[\phi
_{1},U_{1}]$ becomes of the same order order as linear terms, that
is for $A\gtrsim \eta ^{2/3}$. At this point, $m=0$ and $m=2$
components begin to be "fed" nonlinearly by mode coupling terms: the
$m=0$ and $m=2$ modes are nonlinearly driven. However, these mode
coupling terms, quadratic in $A$, do not affect the $m=1$ dynamics
so that one could say that the $m=1$ mode is still linear. At this
stage, it is easy to check that the dominant equations on the $m=1$
component are still the linear ones. This means that the radial
structure of the solution should remain close to the linear one.
Given that, it is possible to include the correction to the linear
theory due to the new location of the critical layer. Because of the
$m=1$ perturbation, the critical layer is not expected to remain
fixed at $r_{s0}$. The real helical magnetic field potential inside
the critical layer (for $\left\vert
r-r_{s0}\right\vert \lesssim \eta ^{1/3}$) is%
\begin{equation}
\psi (r,\alpha ,t)=\frac{1}{2}\psi ^{\prime \prime
}(r_{s0})(r-r_{s0})^{2}+A(t)\psi _{L}(r)\cos \alpha .  \label{after_linear}
\end{equation}%
In writing down the critical layer equations, the instantaneous surface $%
r_{s}(\alpha ,t),$ defined by $\partial _{r}\psi \left[ r_{s}(\alpha ,t)%
\right] =0$, is important as the location where dynamical equations turn
singular and non-ideal effects come into play. As mode couplings do not
affect the second order $m=1$ dynamics, one can keep the linear $m=1$ radial
structure but introduce the corrections due to the motion of the $\partial
_{r}\psi =0$ surface. This yields a differential equation \cite%
{Firpo04,FirpoCoppi03} for the $m=1$ amplitude $A(t)$ valid below the onset
of \textquotedblleft truly nonlinear\textquotedblright\ cubic nonlinearities
on $m=1$. This is given by
\begin{equation}
\frac{dA}{dt}=\gamma (t)A(t)  \label{diff_equa_megal1}
\end{equation}%
with%
\begin{equation}
\frac{\gamma (t)}{\left\langle \frac{\psi _{0}^{\prime \prime }\left[
r_{s}(\alpha ,t)\right] }{r_{s}(\alpha ,t)}\right\rangle _{\alpha }}=\frac{%
\gamma (t=0)}{\left\langle \frac{\psi _{0}^{\prime \prime }\left[
r_{s}(\alpha ,t=0)\right] }{r_{s}(\alpha ,t=0)}\right\rangle _{\alpha }}=%
\frac{\gamma _{L}}{\frac{\psi _{0}^{\prime \prime }\left( r_{s0}\right) }{%
r_{s0}}},  \label{diff_equa_effective}
\end{equation}%
where $\left\langle \cdot \right\rangle _{\alpha }\equiv \left( 2\pi \right)
^{-1}\int_{0}^{2\pi }\cdot d\alpha $ denotes the $m=0$ average. As shown in
Refs. \cite{Firpo04,FirpoCoppi03}, the differential equation (\ref%
{diff_equa_megal1}), with $\gamma (t)$ given by (\ref{diff_equa_effective}),
may actually be approximated by the quadratic expression%
\begin{equation}
\frac{dA}{dt}=\gamma _{L}A(t)-C_{0}A^{2}(t),
\label{diff_ampl_eq_second_order}
\end{equation}%
with
\begin{equation}
C_{0}=\frac{q^{\prime }(r_{s0})+r_{s0}q^{\prime \prime
}(r_{s0})-2r_{s0}q^{\prime }(r_{s0})^{2}}{\sqrt{2}\pi r_{s0}^{2}q^{\prime
}(r_{s0})}.  \label{defi_C0}
\end{equation}%
It may be useful to remind here that the problem has been rendered
dimensionless. The solution of (\ref{diff_ampl_eq_second_order}) is%
\begin{equation}
A(t)=\frac{A_{0}\exp \left( \gamma _{L}t\right) }{1+A_{0}C_{0}/\gamma _{L}%
\left[ \exp \left( \gamma _{L}t\right) -1\right] }
\label{solu_ampl_eq_second_order}
\end{equation}%
that tends to $\gamma _{L}/C_{0}$ as $t\rightarrow \infty $. It is
interesting to note that the curve $A/A_{0}$ has a universal form depending
only on the rescaled time $\gamma _{L}t$ and on the parameter $%
A_{0}C_{0}/\gamma _{L}$ containing the $q$-profile properties at $r_{s0}$.
It is also interesting to note that, at time $t_{i}=\gamma _{L}^{-1}\ln %
\left[ \gamma _{L}/\left( A_{0}C_{0}\right) -1\right] $, $A(t)$
possesses an inflexion point so that, around that time, the
effective behavior of $A$ is approximately algebraic (being linear).
All this assumes that $C_{0}$ is positive. A negative $C_{0}$ would
yield a transient explosive faster-than-exponential behavior.
However, the validity of (\ref{diff_ampl_eq_second_order}) is
limited because some cubic nonlinearities should come into play. For
$S$ not too large, so that the instantaneous second order location
of the critical
layer has some large overlap with the initial linear critical layer at $%
r_{s0}$ around the X-point, it is possible \cite{Firpo04,FirpoCoppi03} that
those cubic terms show up in a spectacular manner. When collecting terms
cubic in $A$, it turns out that in this overlap domain the convective
derivative due to the motion of the critical layer dominates the ordinary
time derivative and equilibrates mode coupling terms yielding, together with
(\ref{diff_ampl_eq_second_order}), an effective amplitude equation of the
form%
\begin{equation}
\frac{dA}{dt}-\gamma _{L}A+C_{0}A^{2}+\frac{c}{\eta }A^{2}\left( \frac{dA}{dt%
}-\gamma _{NL}A\right) =0  \label{total_diff_eq_A}
\end{equation}%
with $c=\mathcal{O}(1)$ some constant and $\gamma _{NL}$ the growth rate
reached by the $m=1$ mode at the onset of cubic terms. This takes place for $%
A\sim \eta ^{1/2}$, so that $\gamma _{NL}\sim \gamma _{L}\eta ^{1/2}\left( 1-%
\frac{C_{0}}{\gamma _{L}}\eta ^{1/2}\right) $.

\begin{figure}[tbp] \centering
\includegraphics[width=10cm]{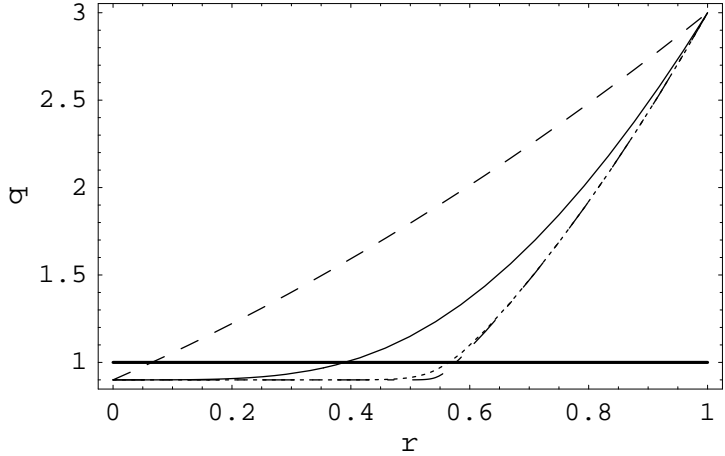}
\caption{Different $q$-profiles of the family $q(r)=q_{0}\left\{
1+r^{2 \protect\lambda}\left[ \left( q_{a}/q_{0}\right)
^{\protect\lambda}-1\right]
\right\} ^{1/\protect\lambda}$ with $q_{0}=0.9$ and $q_{a}=3$ for $\protect%
\lambda =0.6$ (dashed line), $\protect\lambda=2$ (plain), $\protect\lambda%
=10 $ (dots) and $\protect\lambda=30$ (dot-dashed line). The bold
line is the $q=1$ threshold.} \label{fig-family-q}
\end{figure}

\begin{figure}[tbp]
\centering
\includegraphics[width=10cm]{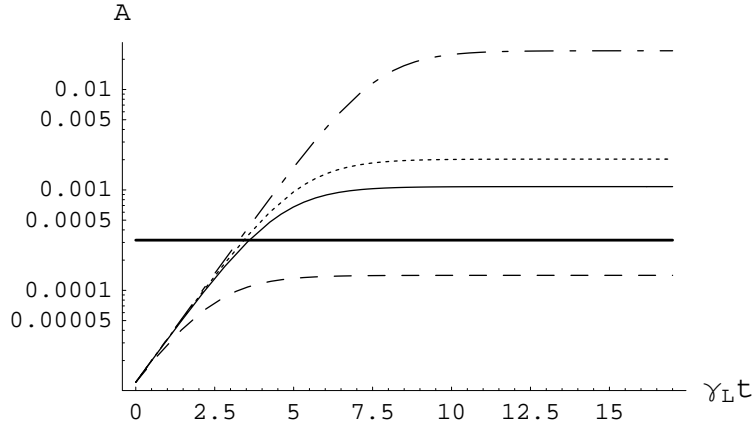}
\caption{Evolution (in linear-log scale) of the amplitude of the
$m=1$ mode, given by Eq. (\ref{solu_ampl_eq_second_order}), as a
function of time normalized to the linear growth rate for the
$q$-profiles displayed in Fig. \protect\ref{fig-family-q} (with the
same plot styles) before the onset of "third order" nonlinear
regime. The bold horizontal line marks the threshold of third order
convective terms for $A\sim S^{-1/2}$ for the value $S=10^{7}$. The
initial amplitude is $A_{0}=1.2\times 10^{-5}$. }
\label{fig-amplitudes}
\end{figure}

Let us finally investigate the influence of the $q$-profile in the onset of
the nonlinear regime. The $q$-profiles $q(r)=q_{0}\left\{ 1+r^{2\lambda }%
\left[ \left( q_{a}/q_{0}\right) ^{\lambda }-1\right] \right\} ^{1/\lambda }$%
, parameterized by $\lambda $, can depict different possible
experimental situations (see Figure \ref{fig-family-q}).
The low$\lambda $ case is consistent with a very peaked current
profile. Such a kind of profile was used for instance by Biskamp in
his 1991's simulations \cite{Biskamp91} of the same system. He
observed a transition from the linear exponential $m=1$ growth
towards an algebraic behavior. On the contrary, the large $\lambda $
case coincides with a low shear situation with a flat current
profile within the $q=1$ radius. This is reminiscent of recent
experimental investigations undertaken e.g. in JET under the
"hybrid" scenario, with a wide area of low magnetic shear and
central safety factor close to and below one. Buratti and coworkers
have reported in this case the wide emergence of "slow sawteeth"
\cite{BurattiEPS04,Gormezano04} where the mode has the same spatial
structure as the kink-like sawtooth precursor with $n=1$ but grows
very slowly and enters the nonlinear regime at the linear growth
rate. Although this case may not strictly correspond to the purely
resistive mode, the present analysis on the onset of nonlinear
effects should be transposable. Figure \ref{fig-amplitudes} may
indeed propose an explanation for these observations. In the case of
a very peaked $q$-profile (e.g. with $\lambda =0.6$), the
integration of $m=1$ amplitude evolution Eq.
(\ref{solu_ampl_eq_second_order}) before the onset of third order
convective effects shows that $A$ should saturate before this third
order threshold $A\sim \eta ^{1/2}$ is reached.
In particular, this may explain Biskamp's observations
\cite{Biskamp91} of a transition to an algebraic stage with no
subsequent nonlinear exponential growth. The $\lambda =2$ case just
corresponds to the $q$-profile taken by Aydemir in Ref.
\cite{Aydemir97}. Here the third order convective stage showing a
second stage of exponential growth can be reached. It is clear from
Fig. \ref{fig-amplitudes} that the nonlinear growth rate $\gamma
_{NL}$, that is the growth rate of the $m=1$ mode when $A$ crosses
the threshold ($A\sim \eta ^{1/2}$) is (slightly) smaller than
$\gamma _{L}$ as $A$ has turned bending. Indeed the simulations of
Ref. \cite{Aydemir97} give the numerical value $\gamma _{NL}\simeq
\gamma _{L}/2$. Yet, for larger values of $\lambda $, corresponding
to a $q$-profile of the kind studied by Buratti et al.
\cite{BurattiEPS04,Gormezano04}, it is clear from Figure
\ref{fig-amplitudes} that $\gamma _{NL}$ would be almost equal to
$\gamma _{L}$: This is an explanation of the fact that, in "slow
sawteeth", the $m=1$ mode enters the nonlinear regime at almost the
linear growth rate.

% \appendix

% \section{}
% \label{}

\ack

The author is greatly indebted to B. Coppi for his advice and explanations
and thanks the organizers and participants of the 9th Plasma Easter Meeting
held in Torino for the nice meeting.

%\bibliographystyle{unsrt}
%\bibliography{bibEFTC}

%\begin{thebibliography}{00}


% notes:

% \bibitem{label2}
% If there is a note, it should come last:
% \bibitem{label3} \note
% \end{subbibitems}

%\bibitem{}

%\end{thebibliography}

\end{document}